\documentclass[pre,twocolumn,aps,eqsecnum]{revtex4}
\usepackage{epsfig}

\begin{document}

\title{Shear-Transformation-Zone Theory of Yielding in Athermal Amorphous Materials}

\author{J.S. Langer}
\affiliation{Department of Physics, University of California, Santa Barbara, CA  93106-9530}

\date{\today}

\begin{abstract}
Yielding transitions in athermal amorphous materials undergoing steady-state shear flow resemble critical phenomena.  Historically, they have been described by the Herschel-Bulkley rheological formula, which implies singular behaviors at yield points. In this paper, I examine this class of phenomena using an elementary version of the thermodynamic shear-transformation-zone (STZ) theory, focusing on the role of the effective disorder temperature, and paying special attention to scaling and dimensional arguments.  I find a wide variety of Herschel-Bulkley-like rheologies but, for fundamental reasons not specific to the STZ theory, conclude that the yielding transition is not truly critical.  In particular, for realistic many-body models with short-range interactions, there is a correlation length that grows rapidly, but ultimately saturates near the yield point. 

\end{abstract}

\maketitle

\section{Introduction}
\label{Intro}

Numerical simulations and analytic approximations imply that yielding transitions in athermal amorphous materials, undergoing steady-state shear flow, resemble critical phenomena.  These transitions are characterized by fluctuating regions of correlated, irreversible, molecular or granular rearrangements, whose sizes grow as the shear rates decrease and the systems approach their yield points.  Recent work on this subject has been influenced by the stochastic model introduced in 1998 by H\'ebraud and Lequeux.\cite{HEBRAUD-LEQUEUX-1998} For example, Bocquet,  Colin and Adjari \cite {BOCQUET-etal-2009}, and Nicolas, Martens and Barrat \cite{BARRAT-etal-14} have studied related models that focus on elastic interactions between localized, stress-driven events. Lemaitre and Caroli \cite{LEMAITRE-CAROLI-09} have studied yielding in zero-temperature, molecular-dynamics simulations of two-dimensional, binary, Lennard-Jones mixtures.  Lin et al. \cite{LINetal-14} and Muller and Wyart \cite{MULLER-WYART-15} have developed new techniques for studying instabilties and scaling behaviors for models of noncrystalline materials near yield points. Common outcomes of these  investigations are versions of the well known Herschel-Bulkley (HB) rheology in which the incremental stress above the yield point is proportional, in many of these cases, to the square root of the plastic shear rate.  This singular behavior, if accurate, would imply that some kind of collective motion is occurring at the transition.

From its inception, the theoretical picture that has come to be known as ``Soft Glassy Rheology'' (SGR) \cite{SGR-SOLLICHetal-1997,SGR-SOLLICH-1998} focused explicitly on Herschel-Bulkley and related rheological behaviors.  In contrast, the shear-transformation-zone (STZ) theory \cite{FL-98,FL-11} was never presented clearly in rheological language, which now seems to me to have been unfortunate.  The STZ theory is an attempt to  describe amorphous plasticity in terms that are closer than SGR to the underlying physics and statistical thermodynamics of these processes.  It is an intrinsically mean-field theory; it does not explicitly describe correlations between localized shear transformations, at least not in its present form.  However, it does describe mean-field collective motion via a thermodynamically defined, effective disorder temperature.  The noise generated by driven plastic deformation disorders the system, raises the effective temperature, and thereby creates new flow defects (STZ's) that, in turn, contribute to the deformation rate.  This nonlinear mechanism provides a framework in which to describe a variety of Herschel-Bulkley-like rheological behaviors. A similar mechanism has been shown to account for the basic features of dislocation-induced plasticity in polycrystalline materials \cite{LBL-Dislocations-10}, where the generation of defects, i.e. dislocations, is a hardening rather than a softening mechanism.  

In fact, HB behavior has appeared implicitly in several earlier STZ-related publications. For example, in \cite{JSL-MANNING-TEFF-07} (LM), Manning and I analyzed data from glass-dynamics simulations by Haxton and Liu \cite{HAXTON-LIU-07} (HL), who saw HB power laws at low temperatures and in a range of relatively high shear rates.  Importantly for present purposes, HL also determined effective temperatures directly by measuring pressure fluctuations. Similar rheological results appear (less clearly) in my papers with Egami \cite{JSL-EGAMI-12} and Lieou \cite{LIEOU-JSL-12}. In all three of these investigations, our primary interest was in the transitions from yielding to viscous behavior as functions of temperature or packing fraction.  We presented our results in the form of log-log plots, and thus did not pay attention to the rheological significance of the low-temperature data, nor did we pay close enough attention to the yielding transition itself. My purpose here is to look harder at the latter aspects of the theory.  I particularly want to emphasize the role of fundamental, dimensional and scaling arguments. It is these general arguments, and not system-specific ones, that lead to the conclusion that yielding in athermal amorphous materials is not a true critical transition. 

\section{STZ Basics}
\label{STZ Basics}

A central feature of the STZ theory is that it treats the density and orientations of localized flow defects as dynamical variables.  The STZ's fluctuate into and out of existence in the environment of an elastic solid.  When present, they undergo irreversible rearrangements and thus produce plastic deformation in response to stresses. These local transitions have been observed directly in simulations and, recently, in a detailed experimental study of a colloidal glass by Jensen, Weitz, and Spaepen. \cite {JENSEN-et-al-2014}. The STZ equations of motion that determine these behaviors are subject to constraints imposed by the first and second laws of thermodynamics.  This theory has been discussed extensively in the literature \cite{FL-11,BL-I-II-III-09} along with its applications to shear banding \cite{MANNINGetal-SHEARBANDS-08}, fracture toughness \cite{RYCROFT-EB-12}, oscillatory viscoelasticity \cite{BL-11}, and the like. In what follows, I briefly summarize the main features of the STZ theory without repeating detailed derivations.  As part of this summary, however, I raise several fundamental issues that I think need special emphasis.

For athermal systems undergoing steady-state flow in pure shear, the basic STZ relation between deviatoric stress $\sigma$ and plastic shear rate $\dot\gamma^{pl}$ has the form
\begin {equation}
\label{q-chi-f}
q \equiv \dot\gamma^{pl}\,\tau_0 = \epsilon_0\,e^{- 1/\chi}\,f(\sigma).
\end{equation}
Here, $\chi$ is the effective temperature expressed in units of a characteristic STZ formation energy; and the Boltzmann factor $e^{-1/\chi}$ is proportional to the density of STZ's. $\epsilon_0$ is a dimensionless constant, roughly of the order of unity.  It is the product of the prefactor of the Boltzmann exponential, i.e. a site density, and the volume of the deformable core of an STZ. $f(\sigma)/\tau_0$ is the stress-dependent STZ transition rate, to be specified below. For simplicity, I consider only pure shear in the $x,\,y$ plane, so that $q = q_{xx}= - q_{yy}$ and $\sigma = \sigma_{xx} = - \sigma_{yy}$. 

Note that I have introduced a time constant $\tau_0$ in Eq.(\ref{q-chi-f}). Even in athermal systems such as those considered here, the internal dynamics must determine a nonzero time scale on which local configurational fluctuations relax and dissipate energy.  By  introducing $\tau_0$, I emphasize that I am not considering athermal, quasi-stationary, numerical models, which exclude any notion of time whatsoever, and therefore do not allow us to distinguish between fast and slow processes as is done implicitly in the definition of $q$ in Eq.(\ref{q-chi-f}). Nor am I considering models with infinitely long ranged interactions, or even jammed granular materials in which force chains may span the system.  $\tau_0$ must be a well defined, few-body relaxation time.  Systems that are driven faster than $1/\tau_0$  behave differently than those that are driven more slowly; thus $q \sim 1$ plays a special role in the following analysis.

Derivations of the Boltzmann factor on the right-hand side of Eq.(\ref{q-chi-f}) appear in STZ papers such as  \cite{FL-11,BL-I-II-III-09}.  As shown there, this elementary statistical formula is valid, not just for equilibrated systems, but also for systems that are driven persistently away from equilibrium.  It follows directly from the requirement that the statistically defined entropy be a non-decreasing function of time. $\chi$ is the derivative of the configurational energy with respect to the configurational entropy (see Sec.~\ref{Teff}); it should not be thought of as a ``noise temperature'' or used, for example, in a Langevin equation -- certainly not without some systematic rationale.      

For present purposes, the rate factor in Eq.(\ref{q-chi-f}) can be written in the form
\begin{equation}
\label{fdef}
f(\sigma) = \cases{{\cal C}(\sigma) \left(1 - {\sigma_y\over |\sigma|}\right)& if $|\sigma| > \sigma_y$; \cr 0 & if $|\sigma| < \sigma_y$.}
\end{equation} 
The yield stress $\sigma_y$ locates an exchange of dynamic stability between non-flowing and flowing states.  The fully time dependent STZ equations of motion (see \cite{FL-11}) describe the behavior of the density of STZ's and their average orientations with respect to the stress.  These equations have stable fixed points for flowing states only if $|\sigma| > \sigma_y$, where the rate at which STZ's are deactivated by making transitions into the forward direction is balanced by the rate at which new STZ's are formed. In the athermal limit, $\sigma_y/|\sigma|$ is the fraction of STZ's that is aligned with the stress; in Eq.(\ref{fdef}), it plays the role of a back stress.  $\sigma_y$ is a system-specific quantity.  It first appears as a factor in the relation between the rate of work done by the external driving force and the strength of the mechanical noise that induces STZ creation and annihilation. (See the discussion of the noise frequancy $\Gamma$ in Sec.\ref{Heterogeneity}.) 

The factor ${\cal C}(\sigma)$ in Eq.(\ref{fdef}) is a linear combination of STZ transition rates, and is a symmetric function of $\sigma$.  In LM, Manning and I wrote this term in the form
\begin{equation}
{\cal C}(\sigma)= \left[1 + \left({\sigma\over \sigma_1}\right)^2\right]^{n/2}.
\end{equation}
Here, I assume that any overall multiplicative constant on the right-hand side of this equation has been incorporated into the rate factor $1/\tau_0$.  For larger stresses, of the order of or greater than some $\sigma_1$, Manning and I assumed that this rate would grow as the $n$'th power of the stress.  

There are numerous possible rationales for choosing $n$.  For example, $n = 1$ would be appropriate for a colloidal suspension in which particle motions are subject to linear viscous drag. A different possibility, for systems in which particles interact via short-ranged repulsive forces, is to invoke Bagnold scaling.  As conjectured in LM, if there are no natural stress scales in the system other than the rate-independent $\sigma_y$ and $\sigma_1$, then dimensional analysis might require that the dynamic stress $\sigma$ be proportional to an acceleration, i.e. an inverse time squared, and thus be proportional to the square of a dynamic rate.  Conversely, the rate ${\cal C}(\sigma)/\tau_0$ would then be proportional to the square root of the stress.  This result, by itself, is inconsistent with HB rheology.  Nevertheless, the Bagnold assumption with $n = 1/2$ fits the Haxton-Liu data quite nicely because, as shown in LM, the effective temperature $\chi$ and the corresponding STZ density are functions of $q = \dot\gamma^{pl}\,\tau_0$, which brings the time scale $\tau_0$ back into the theory.  This Bagnold analysis seems to me to be plausible but not compelling in the present context. 

\section{Effective Disorder Temperature}
\label{Teff}

The equation of motion for the effective temperature $\chi$ is a statement of the first law of thermodynamics, i.e. energy conservation.  Let $U$ denote the configurational internal energy as a function of the configurational entropy $S$.  Then $\chi = \partial U/\partial S$; and the first law takes the form
\begin{equation}
\tau_0\,\chi\,\dot S \approx \tau_0\,V\,c_{e\!f\!f}\,\dot\chi = 2\, V\,q\,\sigma - Q,
\end{equation}
where $\chi\,\dot S$ is the rate of change of $U$, $c_{e\!f\!f}$ is an effective specific heat, $2\,q\,\sigma/\tau_0$ is the rate at which work is done on the system per unit volume, $V$ is the volume, and $Q/\tau_0$ is the rate at which heat is dissipated. The second law requires that $Q = V\,\kappa\,(\chi - \theta)$, where $\kappa$ is a non-negative heat transfer coefficient, and $\theta$ is the ambient temperature expressed in the same units as those used for $\chi$.  For the athermal systems considered here, $\theta \approx 0$. Thus, the steady-state equation of motion for $\chi$ can be written in the form
\begin{equation}
\label{dotchi}
\tau_0\,\dot\chi \approx {2\, q\,\sigma\over c_{e\!f\!f}} - \kappa \chi \equiv {2\, q\,\sigma\over c_{e\!f\!f}}\,\left[1 - {\chi\over \hat\chi(q)}\right] = 0.
\end{equation}
With these assumptions, I have defined $\chi \equiv \hat\chi(q)$ in steady state.  Note that the underlying physics is thermal transport, which somehow must be encoded in $\hat\chi(q)$.  

To evaluate $\hat\chi(q)$, it is useful to think about the analogous relation between ordinary temperature and relaxation rates in glasses.  Note first that, if we shear or otherwise ``stir'' a system more slowly than any of its internal relaxation rates, i.e. set $q \ll 1$, then this system must ultimately reach a $q$-independent effective temperature, say $\chi_0$.  Shearing an amorphous material necessarily involves rearranging the particles, i.e. forcing them to move around each other to accommodate the shear deformation.  As we shear slowly, we bring clusters of particles into unstable configurations, from which they relax relatively quickly into new stable  positions. These persistent rearrangements produce a fluctuating steady state of disorder, i.e. an effective temperature $\chi$, that is independent of the shear rate so long as that rate is slow enough.  Achieving the steady-state value of $\chi$ requires only that the accumulated shear strain be large enough, perhaps of the order of unity, so that a statistically significant number of rearrangements occurs. This argument should be valid whether or not the STZ model of those rearrangements is accurate.  It is clearly consistent with the HL molecular dynamics simulations shown  in Fig. 2.

This rationale for asserting that $\chi \to \chi_0 > 0$ in the limit $q \to 0$ is crucial for understanding the fundamental distinction between realistic many-body systems and some of the models that have been proposed to describe yielding, e.g. refs. \cite{HEBRAUD-LEQUEUX-1998,BOCQUET-etal-2009,BARRAT-etal-14} or the recent AQS study by Salerno and Robbins \cite{SALERNO-ROBBINS-2013}.  The authors of \cite{SALERNO-ROBBINS-2013} cannot compare the imposed strain rate with dynamic relaxation rates, as is necessary in my argument about $\chi_0$.  They also have no way to  determine whether extended cascades of events might be affected -– perhaps limited -– by other events that can occur simultaneously.  Those questions are also well beyond the scope of the simpler models described in refs. \cite{HEBRAUD-LEQUEUX-1998,BOCQUET-etal-2009,BARRAT-etal-14}.  This fundamental observation leads me to conclude that, for purposes of studying yielding transitions, these models are not in the same universality class as realistic many-body systems.   

The glassy analog of this behavior is that the configurational relaxation time $\tau_{\alpha}$, like $1/q$, diverges (or  becomes  unmeasurably large) as the system approaches an ideal glass transition. Conversely, the temperature varies increasingly slowly as $\tau_{\alpha}$ goes to infinity. Like the glass transition temperature, $\chi_0$ is a system-specific parameter.  We will see in Sec.~\ref{HB} that its nonzero value determines the nature of the yielding transition. 

At the other end of the range of shear rates, the glass analogy implies Arrhenius behavior.  Sufficiently far above the glass temperature, the configurational relaxation rate $\tau_{\alpha}^{-1}$ (or the inverse viscosity, or the diffusion constant, or the like) appears experimentally to be controlled by a thermally activated process with a temperature-independent energy barrier.  Similarly, for values of $q$ not too small, the Haxton-Liu measurements of the effective temperature are well described by a relation of the form $q \approx q_0\,e^{-A/\chi}$, where $A$ is an activation energy expressed here, like $\chi$, in units of the STZ formation energy.  To interpolate between these two limiting behaviors in \cite{JSL-MANNING-TEFF-07,JSL-EGAMI-12,LIEOU-JSL-12}, my coauthors and I have postulated a modified Vogel-Fulcher-Tamann (VFT) formula:
\begin{equation}
\label{hatchi}
q(\hat\chi) = q_0\,\exp\,\left[-{A\over \hat\chi} - \alpha_{eff}(\hat\chi)\right],
\end{equation}
where
\begin{equation}
\label{VFT}
\alpha_{eff}(\hat\chi) = \left({\chi_1\over \hat\chi - \chi_0}\right)\,\exp\,\left[-\,b\,{\hat\chi - \chi_0\over \chi_A-\chi_0}\right].
\end{equation}
The first term on the right-hand side of Eq.(\ref{VFT}) is the conventional VFT divergence at $\hat\chi = \chi_0$.  Here it is cut off in $\hat\chi$ by an exponential factor so that the behavior of $q(\hat\chi)$ in Eq.(\ref{hatchi}) is dominated by the Arrhenius term for $\hat\chi - \chi_0 > \chi_A - \chi_0$.  This interpolation formula has no special physical significance that I am aware of and, so far, its details have not seemed crucial for interpreting experimental or computational data. On the other hand, the parameter $A$ plays a special role in what follows. 

In interpreting Eq.(\ref{hatchi}), note that the STZ theory  has an absolute upper limit of validity where $\hat\chi \to \infty$.  At such large values of $\hat\chi$, the density of STZ's becomes large, and the theory is no longer consistent with a model of a solid containing a dilute population of flow defects.  A natural way to formulate the theory, then, is to choose $q_0 =1$ so that the maximum theoretical shear rate is $\dot\gamma^{pl}_{max} = 1/\tau_0$.  Above this rate, the system has insufficient time to relax between irreversible rearrangements; i.e., it behaves like a fluid.  Adopting this convention, however, means that we need to use a physically realistic estimate of $\tau_0$.
\begin{figure}[here]
\centering \epsfig{width=.5\textwidth,file=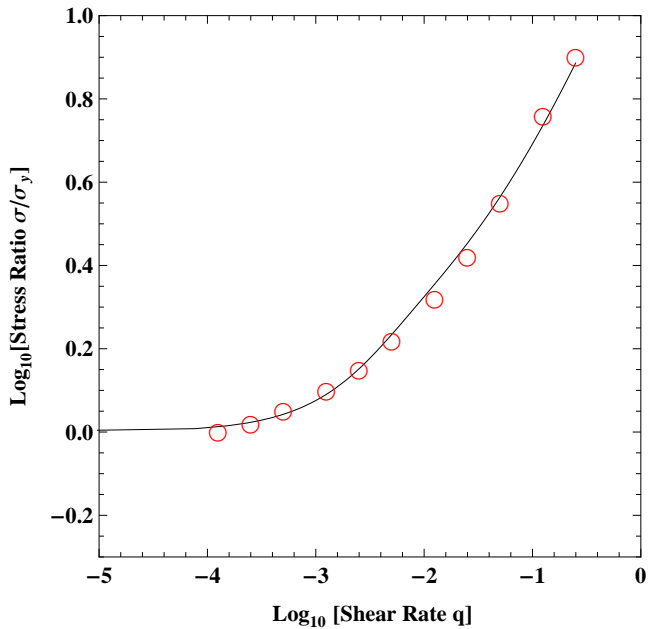} \caption{Log-Log plot of stress $\sigma$ in units of the yield stress $\sigma_y$ as a function of dimensionless shear rate $q$.  The data points are from HL, with rescaled values of $q$ as described in the text. } \label{HBFig1}
\end{figure}

\section{Herschel-Bulkley Behavior}
\label{HB}

At large stresses and shear rates, the preceding set of equations immediately produces a generalized Herschel-Bulkley relation.  In this limit,
\begin{equation}
q \approx \epsilon_0\, e^{-1/\chi} \left({\sigma\over \sigma_1}\right)^n \approx q_0\,e^{-A/\chi}.
\end{equation}
Eliminating $\chi$, we find
\begin{equation}
\label{HBlimit}
\sigma \approx \sigma_1\,\left({q_0\over \epsilon_0}\right)^{1/n} \left({q\over q_0}\right)^{\beta};~~~~~\beta = {A-1\over n\,A}.
\end{equation} 
Then, for example, the Bagnold choice $n = 1/2$, and $A = 4/3$, produces the conventional HB power law, $\beta = 1/2$.  However, with different choices of parameters, this STZ-based theory of athermal yielding produces a wide range of rheological behaviors, well beyond the simple square-root law.

\begin{figure}[here]
\centering \epsfig{width=.45\textwidth,file=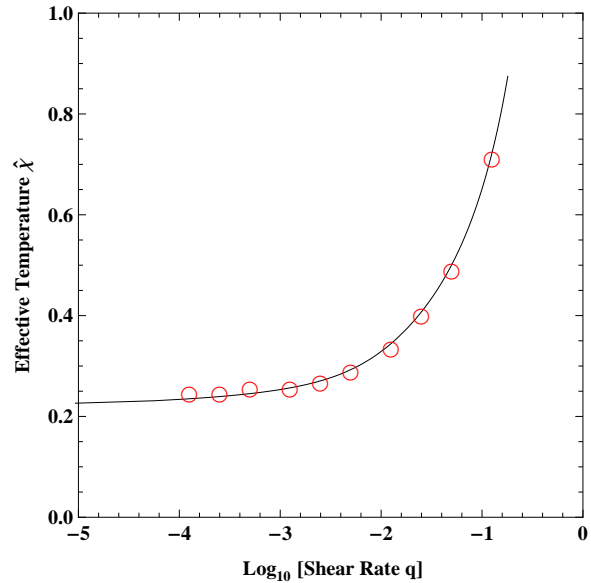} \caption{Semi-log plot of the steady-state effective temperature $\hat\chi$ as a function of dimensionless shear rate $q$.  The data points are from HL, with rescaled values of $q$ as described in the text. } \label{HBFig2}
\end{figure}

To explore some of these possibilities, I start by reanalyzing the low temperature HL data. I have made two changes from the earlier LM analysis.  First, instead of using the microscopic time scale adopted by HL, I have deduced a presumably more realistic $\tau_0$ by setting $q_0 = 1$. The resulting $\tau_0$ is about a factor $12$ larger than the original HL value, but is still a microscopic time scale.  Second, in our previous attempt to fit all the HL data, including at high temperatures, with as few parameters as possible, Manning and I used what I now think was an unrealistically small value of $\sigma_1$, and thus needed to use a correspondingly small value of $\epsilon_0$.  In simple athermal systems, however, there is really only one physically meaningful stress scale, which is set by the shear modulus $\mu$.  Both $\sigma_y$ and $\sigma_1$ should be very roughly of the order of $\mu$.  Therefore, I have set $\sigma_1 = \sigma_y$, and have measured the dynamic stress $\sigma$ in units of $\sigma_y$.  Then I have adjusted $\epsilon_0$ to fit the data, finding $\epsilon_0 \cong 0.26$ in accord with my expectation that this parameter should be roughly equal to unity.  Other parameters, as given in LM, are: $A = 1.5$ (so that $\beta = 2/3$ as observed), $\chi_0 = 0.2$ (directly observed as shown in Fig.~\ref{HBFig2}), $\chi_1 = 0.26$, $\chi_A = 0.3$, and $b= 3$.

\begin{figure}[here]
\centering \epsfig{width=.45\textwidth,file=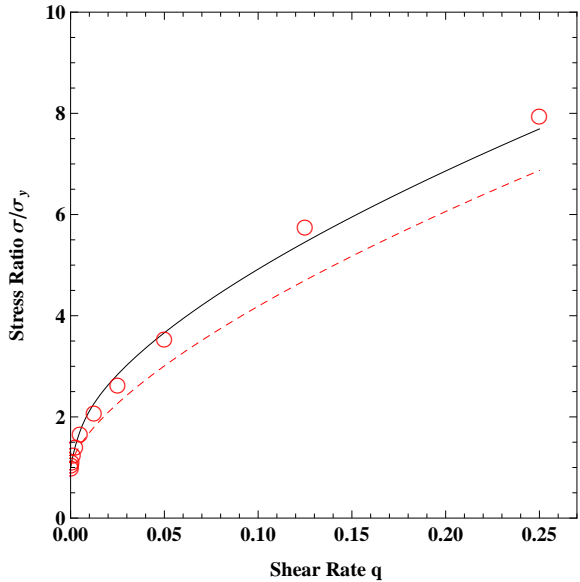} \caption{Stress $\sigma$ in units of the yield stress $\sigma_y$ as a function of dimensionless shear rate $q$.  The data points are from HL, with rescaled values of $q$ as described in the text. The dashed line is the approximation given in Eq.(\ref{HBlimit}) displaced upward by one unit of the yield stress.} \label{HBFig3}
\end{figure}

Figures \ref{HBFig1} and \ref{HBFig2} show the HL data in their original form as a log-log plot of $\sigma/\sigma_y$ versus shear rate $q$, and a semilog plot of the effective temperature $\hat\chi(q)$, along with the theoretical curves evaluated using the full equations given above.  In Fig.~\ref{HBFig2}, note that the upper limit of the data at $q \cong 0.25$ occurs at $\hat\chi \cong 0.7$, which should be about at the limit of validity of the STZ theory, as intended. Figure \ref{HBFig3} is a direct plot of the data in which the HB form is easily visible, with the crossover to the $q \ll 1$ behavior squeezed into a small part of the graph near $q = 0$.  For comparison, the dashed curve shows the approximation in Eq.(\ref{HBlimit}) displaced upward by one unit of the yield stress. As can just be seen in both the data and the theory, the stress rises linearly above the yield point: 
\begin{equation}
\label{onset}
\sigma - \sigma_y \approx q\,{\sigma_y\over\epsilon_0}\,e^{1/\chi_0}.
\end{equation}

\noindent For the chosen parameters, this formula means that the initial slope of the graph in Fig.~\ref{HBFig3} is large but finite, approximately $570$.  The rapid growth of $\chi$ and the corresponding growth of the STZ density cause the material to soften, with the result that the curve bends over into  power-law behavior at small values of $q$.

\begin{figure}[here]
\centering \epsfig{width=.45\textwidth,file=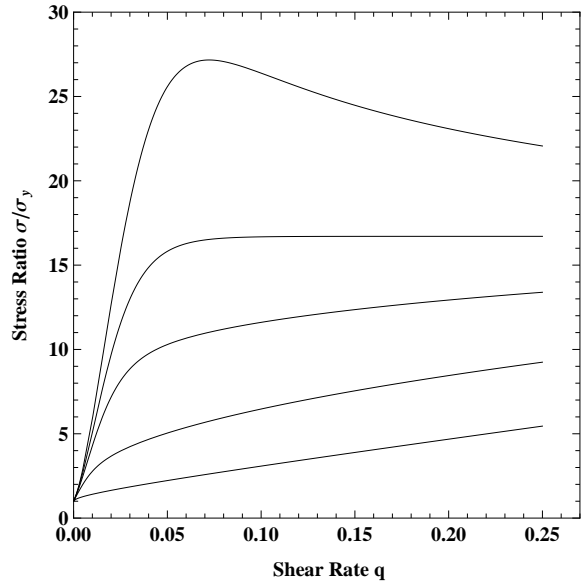} \caption{Stress $\sigma$ in units of the yield stress $\sigma_y$ as a function of dimensionless shear rate $q$, for five different values of the dimensionless activation energy $A$. From bottom to top,  $A = 2.0,~ 1.33,~1.1,~ 1.0,~{\rm and}~ 0.9$.  The corresponding values of the HB exponent are $\beta = 1.0,~0.5,~0.18,~0,~{\rm and}~-0.22$.  } \label{HBFig4}
\end{figure}

To illustrate the variety of rheological behaviors that emerge from this theory, I show in Fig.~\ref{HBFig4} a set of five different curves of stress versus shear rate, using all but one of the same parameters that were used to fit the HL data shown in Figs.~\ref{HBFig1} - \ref{HBFig3}.  The exception is that I have chosen a sequence of different values of the dimensionless activation energy $A$ that controls the dissipation rate in Eq.(\ref{dotchi}).  From bottom to top in the figure, these values are $A = 2.0,~ 1.33,~1.1,~ 1.0,~{\rm and}~ 0.9$. The corresponding values of the HB exponent are $\beta = 1.0,~0.5,~0.18,~0,~{\rm and}~-0.22$.  The second curve, with $\beta =  0.5$, is the square-root law that appears in the fluctuation-theory literature, e.g. \cite{HEBRAUD-LEQUEUX-1998,BARRAT-etal-14, LEMAITRE-CAROLI-09}; and the third and fourth curves, with small values of $\beta$, are included just to show the transition between stable and unstable rheologies.

The more interesting cases shown in Fig.~\ref{HBFig4} are the first and the fifth. The first, at the bottom with $\beta = 1$, illustrates one of the ways in which this theory can produce a conventional, linear, Bingham rheology.  By far the most realistic way for the linear law to be observed, however, is when $\tau_0$ is a microscopic time, perhaps a few molecular vibration periods, so that shear rates with $q \sim 1$ are out of the range of most experiments.  Then, Eq.(\ref{onset}), with a constant $\chi_0$, is accurate throughout an observable range of shear rates, $q \ll 1$.  This is what happened, for example, in earlier analyses of plasticity in metallic glasses (e.g. \cite{JSL08}), where $\chi_0$ was more nearly of the order of $0.1$ instead of $0.2$.  Yet another possibility, still within the scope of this theory, is when materials are intrinsically disordered and soft, i.e., when $\chi_0$ is large, so that the initial slope of $\sigma(q)$ in Eq.(\ref{onset}) is small and remains linear at observably large shear rates.

The fifth interesting case in Fig.~\ref{HBFig4}, at the top with $\beta = - 0.22$, exhibits shear-rate weakening above $q \cong 0.07$, and therefore must be dynamically unstable.  Daub and Carlson \cite {DAUB-CARLSON STICK-SLIP 2009} have  examined a model of this kind in detail, and have shown that it produces both shear-banding and stick-slip instabilies. More recently, Lieou et al. \cite{LIEOU et al PRE 2014 Angular Grains} have applied this kind of theory -- but with $\beta > 0$ -- to granular materials of the kind that occur in fault gouge.  They have added terms to $\kappa$ in Eq.(\ref{dotchi}) to account for shear-rate dependent frictional dissipation due to collisions beween angular grains, and also to account for dissipation induced by tapping or other vibrational perturbations.  In this generalization of the theory, the effective temperature can become a non-monotonic function of $q$, and the volume may also vary non-monotonically. But these modifications are beyond the scope of the present discussion, because they implicitly introduce new  intrinsic time scales analogous to $\tau_0$.  

\begin{figure}[here]
\centering \epsfig{width=.45\textwidth,file=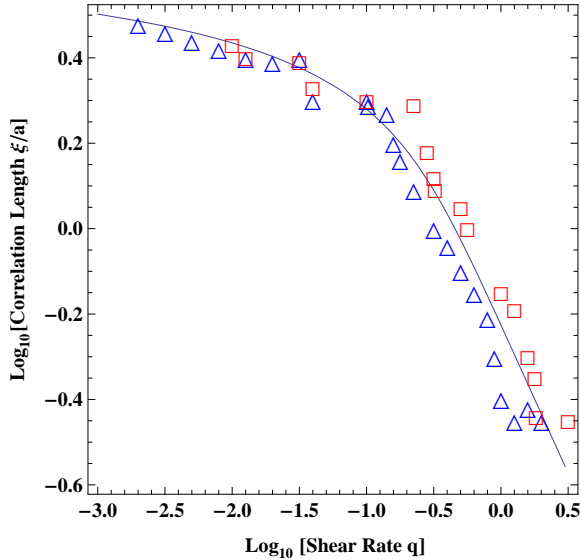} \caption{(Color online) Correlation length $\xi/a$ as a function of dimensionless shear rate $q$. The two sets of experimental data (blue triangles and red squares) are for the two larger forcings shown in Fig.5 of Jop et al  \cite{JOP-etal-2012}, and the theoretical scales of length $a$ and shear rate $q$ are chosen here to be consistent with that data.} \label{HBFig5}
\end{figure}

\section{Spatial Heterogeneities}
\label{Heterogeneity}

As a first step in studying spatial variations in this theory, I have linearized the equation of motion for $\chi$, Eq.(\ref{dotchi}), in the space and time dependent variable $\chi' = \chi - \hat\chi(q)$, and have added a diffusion term:
\begin{equation}
\label{dotchi2}
\tau_0\,\dot\chi' \approx -\,{2\,\sigma\,q\over c_{e\!f\!f}}\,{\chi'\over \hat\chi(q)} + {\cal D}\,\nabla^2 {\chi'\over \hat\chi(q)}.
\end{equation} 
Because $\hat\chi(q)$ diverges when $q \to q_0$, it is best to linearize in $\chi'/\hat\chi(q)$, and to define the diffusion coefficient ${\cal D}$ accordingly. Diffusion terms of this kind have appeared previously in the STZ literature, e.g. in \cite{MANNINGetal-SHEARBANDS-08,DAUB-CARLSON STICK-SLIP 2009}; but it seems to me that ${\cal D}$ has never been given the attention that it needs.  ${\cal D}$ must  have the form $\Gamma\,\ell^2$, where $\Gamma/\tau_0$ is the noise-generated attempt frequency for activation and annihilation of STZ's, and $\ell$ is an elementary diffusion length, which I argue must be the average spacing between STZ's.  This is in contrast to my argument in \cite{JSL-PRE-2012}, where I considered self diffusion of particles, and set $\ell$ equal to the size of the STZ core.  Here, I am considering diffusion of energy or perhaps, in the spirit of \cite{BOCQUET-etal-2009, BARRAT-etal-14}, stress fluctuations; thus, the spacing between STZ's seems appropriate.  

To implement these ideas, I use primarily dimensional analysis.  The only scalar quantity in this system with dimensions of rate is the power per unit volume, $\sigma\,q/\tau_0$.  To convert this quantity into a frequency, divide by a stress times a volume (an energy), say, $\sigma_0\,v_0$, and multiply by the only relevant volume, i.e. the volume per STZ, $v_0\,e^{1/\chi}$.  The factors $v_0$ cancel, and subsequent athermal STZ analysis reveals that $\sigma_0$ is equal to the yield stress $\sigma_y$.  Thus, $\Gamma = 2\,q\,(\sigma/\sigma_y)\,e^{1/\chi}$. Similarly, for dimension $d$, $\ell^d = v_0\,e^{1/\chi}$.  The result is
\begin{equation}
{\cal D} = 2\,a^2\,q\,{\sigma\over \sigma_y}\,e^{(1+2/d)/\chi},
\end{equation}
where $v_0 = a^d$ defines a characteristic interparticle length scale $a$.

Static solutions of Eq.(\ref{dotchi2}) have the form $e^{-r/\xi}$, where   
\begin{equation}
\label{xisquared}
\xi^2 = {c_{e\!f\!f}\,{\cal D}\over 2\,\sigma\,q}=  {c_{e\!f\!f}\,a^2\over \sigma_y}\,\exp\,\left[{1+ 2/d\over \hat\chi(q)}\right].
\end{equation}
In the large-$q$ limit where $q \approx q_0\,e^{-A/\hat\chi}$, Eq.(\ref{xisquared}) becomes
\begin{equation}
\label{xi-q}
\xi^2 \approx {c_{e\!f\!f}\,a^2\over \sigma_y}\,\left({q_0\over q}\right)^{2\nu}; ~~~ \nu = {1 + 2/d\over 2\,A}\,.
\end{equation}
This growing length scale saturates near the yielding transition, where $q \to 0$, and $\hat\chi \to \chi_0$.  

The full function $\xi(q)$ is shown in Fig.~\ref{HBFig5} in comparison with data taken from Fig.~5 of Jop et al. \cite{JOP-etal-2012}.  The data shown here were obtained by measuring velocity-velocity correlations in a dense emulsion flowing under pressure through a gap between parallel glass plates.  Correlations between velocity fluctuations perpendicular to the flow were measured as functions of two-point separations parallel to the flow, at fixed distances from the center of the gap and, thus, at fixed shear rates.  In computing the theoretical curve, I have used $q_0 = 18$ in order to match the scale of shear rates $\dot\gamma$ reported by those authors; and I have set $\sqrt{c_{e\!f\!f}/\sigma_y} = 0.08$ in Eq.(\ref{xisquared}), again to match the units of length used in  \cite{JOP-etal-2012}.  To optimize the fit, I chose $A = 1.2$, which means that the HB exponent is $\beta = 1/3$, and $\nu \cong 0.69$ in Eq.(\ref{xi-q}).  Otherwise, the STZ parameters ($\chi_0,~\chi_1,~ \chi_A,~{\rm and}~ b$) are the same as those used in fitting the HL data in Figs. \ref{HBFig1} - \ref{HBFig3}.  This is the best I can do in the absence of more detailed information about the rheological properties of this particular emulsion.

I emphasize that I have not attempted here to construct a complete theory of the experimental observations in \cite{JOP-etal-2012}, even within the assumptions that led to Eq.(\ref{xisquared}).  To do that, I would have had to couple the equation of motion for $\chi$, Eq.(\ref{dotchi2}), to a position-dependent and fully tensorial version of the STZ flow law, Eq.(\ref{q-chi-f}).  More importantly, I would have had to introduce a nonlocal version of that flow law in which the strain rate on the left-hand side is an exponentially weighted average of the driving term on the right-hand side within a neighborhood of size $\xi$. In other words, I would have had to develop an effective-temperature version of the nonlocal rheology described recently by Hennan and Kamrin \cite{HENNAN-KAMRIN PNAS 2013, HENNAN-KAMRIN PRL 2014} in their theory of dense granular flows. Their analysis must be closely related to the one presented here; it should be useful to explore that relationship.

A nonlocal analysis should not make qualitative changes in the results shown in Fig.~\ref{HBFig5}, which relate to measurements at fixed distances away from the center of the gap where shear rates are constant and nonzero. However, a fully nonlocal analysis would be essential for computing the velocity profile across the flow, as measured by Jop et al. \cite{JOP-etal-2012}. We know by symmetry that, in plug flow of this kind, both the shear rate and the shear stress change sign from one side of the gap to the other. Therefore, there is a region near the center of the gap in which $|\sigma| < \sigma_y$, where a local relation of the kind shown in Eq.(\ref{q-chi-f}) cannot be valid.  A nonlocal calculation of the velocity profile is reported in  \cite{JOP-etal-2012}, where it is used to estimate a length scale whose slow shear-rate dependence may be consistent with $\nu = 1/4$ as predicted in \cite{BOCQUET-etal-2009}.  It remains to be seen whether that behavior can be distinguished from the saturation effect predicted here.

\section{Remarks and Questions}

A principal conclusion of this analysis is that the thermodynamic STZ theory, like SGR, predicts a wide range of Herschel-Bulkley-like rheological behaviors, consistent with a long history of experimental observations.  The surprising result is that -- largely due to the properties of the effective disorder temperature $\chi$ -- the yielding transition in this theory is non-critical.  Immediately above the transition, the stress rises linearly in the strain rate, and the correlation length remains non-diverging.  This result contradicts almost all recent theories of yielding in amorphous materials.  I have argued here that this disagreement occurs because those recent theories are based on models or simulation schemes that do not accurately describe the dynamics of realistic many-body systems.  

I emphasize that this predicted deviation from criticality at the yielding transition is not an artifact of any phenomenological assumption, or of any special feature of the STZ theory.  On the contrary, it emerges from first-principles.  As argued in Sec.~\ref{Teff}, the state of disorder at small dimensionless shear rate $q$ must be independent of $q$, i.e. $\chi \to \chi_0 > 0$.  This nonzero degree of disorder means that, in steady-state deformation, there must be a constant density of some kind of flow defects and, therefore, a linear relation between $q$ and the stress increment just above the yield point. This theoretical conclusion is supported by a close look at the numerical simulations shown in Fig.~\ref{HBFig3}, and by the experimental data shown in Fig.~\ref{HBFig5}.  

On the other hand, I emphasize that this unexpected result does not in any way invalidate the evidence that yielding transitions are generally accompanied by large stress fluctuations and cascades of extended, correlated events.  There are two complementary kinds of questions that need to be asked in this regard.  First, the existing theories have been extremely useful in identifying the physical mechanisms that underly yielding dynamics.  In what ways are they accurate?  Where, precisely, might they be missing essential ingredients?  Second, the mean-field STZ theory provides only an approximate description of the interactions between spatially separated yielding events.  In what ways might it be modified in order to be more accurate?  Does it need to be so modified?   

Theoretical evidence in favor of criticality at yielding transitions has come from stochastic models of the kind originated by H\'ebraux and Lequeux \cite{HEBRAUD-LEQUEUX-1998} and extended in \cite{BOCQUET-etal-2009, BARRAT-etal-14}.  As argued in \cite{LINetal-14} and \cite{MULLER-WYART-15}, the yielding mechanism looks as if it should be in the same universality class as the depinning mechanism that produces, for example, broad distributions of slipping events on earthquake faults.  Renormalization-group analyses of depinning models have produced scaling relations similar to the Herschel-Buckley law.  (For example, see Fisher et. al. \cite{Fisher-etal-1997}.) How precise are these analogies?

The H\'ebraux and Lequeux model produces a Herschel-Bulkley exponent of exactly $1/2$ in the limit of zero strain rate.  Even if we assume that some limiting approximation is being made that eliminates the saturation effect (perhaps by implicitly setting $\chi_0 = 0$), the question remains: why is $\beta = 1/2$ in this class of models? How might it be modified to produce other values of $\beta$? We know that other HB exponents do occur, e.g. in the Haxton-Liu results discussed here, and in a wide variety of rheological situations.  According to Eq.~(\ref{HBlimit}), this exponent is determined by the activation energy $A$, which controls the heat flow. Is there any analogous physics in \cite{HEBRAUD-LEQUEUX-1998}?  Conversely, we must ask: What physics determines $A$ in the STZ theory? Under what physical circumstances might we find unstable dynamics with $A < 1$?

What are the corresponding strengths and limitations of the thermodynamic STZ theory?  We know from recent experience  that the STZ equations of motion, when generalized for variations in space and time and coupled to the equations of motion for elastic fields, constitute a self-contained dynamical theory of amorphous rheology.  The theory does a good job of describing phenomena such as shear banding instabilities \cite{MANNINGetal-SHEARBANDS-08}, brittle and ductile failure in the neighborhoods of crack tips \cite{RYCROFT-EB-12}, stick-slip behavior of granular materials in earthquake faults \cite{DAUB-CARLSON STICK-SLIP 2009,LIEOU et al PRE 2014 Angular Grains}, and the like.  In analogy to the well known equations of motion for fluids, it seems reasonable to expect that these rheological equations of motion, when solved for large systems coupled to external forces and boundary conditions, will predict chaotic behaviors with heterogeneous deformations and local failures on many different length and time scales.  Will the results of such calculations be consistent with those based on the depinning analogy?  Or with experimental observations?

A related question is whether the deterministic nature of the rheological theory presented here might be inadequate for describing broad distributions of event sizes.  The scaling analysis in \cite{Fisher-etal-1997} starts with the assumption of a quenched random pinning force.  H\'ebraux and Lequeux couch their analysis in terms of probability distributions over the values of local stresses.  Is that kind of analysis essential?  Or might the observed behaviors be results of deterministic chaos, as was the case in our earlier slider-block studies of earthquake dynamics? \cite{EARTHQUAKES-RMP-1994}  Large-scale numerical solutions of the present rheological equations of motion might help answer such questions. 

\begin{acknowledgments}

This paper was motivated by discussions at the program on ``Avalanches, Intermittency, and Nonlinear Response in Far-from-Equilibrium Solids'' at the Kavli Institute for Theoretical Physics, University of California, Santa Barbara, September - December 2014.  I am especially grateful to Jean-Louis Barrat for his helpful remarks. This research  was supported in part by the U.S. Department of Energy, Office of Basic Energy Sciences, Materials Science and Engineering Division, DE-AC05-00OR-22725, through a subcontract from Oak Ridge National Laboratory.

\end{acknowledgments}


\begin{thebibliography}{99}

\bibitem{HEBRAUD-LEQUEUX-1998} P. H\'ebraud and F. Lequeux, Phys. Rev. Lett. {\bf 81}, 2934 (1998).

\bibitem{BOCQUET-etal-2009} L. Bocquet, A. Colin and A. Adjari, Phys. Rev. Lett. {\bf 103}, 036001 (2009).

\bibitem{BARRAT-etal-14} A. Nicolas, K. Martens and J.-L. Barrat, EPL {\bf 107}, 44003 (2014).

\bibitem{LEMAITRE-CAROLI-09} A. Lemaitre and C. Caroli, Phys. Rev. Lett. {\bf 103}, 065501 (2009).

\bibitem{LINetal-14} J. Lin, E. Lerner, A. Rosso, and M. Wyart, Proc. Nat'l. Acad. Sci. {\bf 111}, 14382 (2014).

\bibitem{MULLER-WYART-15} M. Muller and M. Wyart, Annu. Rev. Condens. Matter Phys. {\bf 6}, 177 (2015).

\bibitem{SGR-SOLLICHetal-1997} P. Sollich, F. Lequeux, P. H\'ebraud and M. Cates, Phys. Rev. Lett. {\bf 78}, 2020 (1997).

\bibitem{SGR-SOLLICH-1998} P. Sollich, Phys. Rev. E {\bf 58}, 738 (1998).

\bibitem{FL-98} M. L. Falk and J. S. Langer, Phys. Rev. E, {\bf 57}, 7192 (1998).

\bibitem{FL-11} M. L. Falk and J. S. Langer, Annu. Rev. Condens. Matter Phys. {\bf 2}, 353 (2011).

\bibitem{LBL-Dislocations-10} J.S. Langer, E. Bouchbinder and T. Lookman, Acta Mat. {\bf 58}, 3718 (2010).

\bibitem{JSL-MANNING-TEFF-07} J.S. Langer and M.L. Manning, Phys. Rev. E {\bf 76}, 056107 (2007).

\bibitem{HAXTON-LIU-07} T. Haxton and A. J. Liu, Phys. Rev. Lett. {\bf 99}, 195701 (2007).

\bibitem{JSL-EGAMI-12} J.S. Langer and T. Egami, Phys. Rev. E {\bf 86}, 011502 (2012).

\bibitem{LIEOU-JSL-12} C.K.C. Lieou and J.S. Langer, Phys. Rev. E, {\bf 85}, 061308 (2012).

\bibitem{JENSEN-et-al-2014} K. Jensen, D. Weitz, and F. Spaepen, Phys. Rev. E {\bf 90}, 042305 (2014).

\bibitem{BL-I-II-III-09} E. Bouchbinder and J. S. Langer, Phys. Rev. E {\bf 80}, 031131, 031132 and 031133 (2009).

\bibitem{MANNINGetal-SHEARBANDS-08} M. L. Manning, J. S. Langer and J. M. Carlson, Phys. Rev. E {\bf 76}, 056106 (2007).

\bibitem{RYCROFT-EB-12} C.H. Rycroft and E. Bouchbinder, Phys. Rev. Lett. {\bf 109}, 194301 (2012).

\bibitem{BL-11} E. Bouchbinder and J.S. Langer, Phys. Rev. Lett. {\bf 106}, 148301 (2011); Phys. Rev. E {\bf 83}, 061503 (2011).

\bibitem{SALERNO-ROBBINS-2013} K.M. Salerno and M. Robbins, Phys. Rev. E {\bf 88}, 062206 (2013)

\bibitem{JSL08} J.S. Langer, Phys. Rev. E {\bf 77}, 021502 (2008). 

\bibitem{DAUB-CARLSON STICK-SLIP 2009} E. Daub and J.M. Carlson, Phys. Rev. E {\bf 80}, 066113 (2009).

\bibitem{LIEOU et al PRE 2014 Angular Grains} C.Lieou, A. Elbanna, J.S. Langer, and J.M. Carlson, Phys. Rev. E {\bf 90}, 032204 (2014).

\bibitem{JSL-PRE-2012} J.S. Langer, Phys. Rev. E {\bf 85}, 051507 (2012).

\bibitem{JOP-etal-2012} P. Jop, V. Mansard, P. Chaudhuri, L. Bocquet, and A. Colin, Phys. Rev. Lett. {\bf 108}, 148301 (2012).


\bibitem{HENNAN-KAMRIN PNAS 2013} D. Hennan and K. Kamrin, PNAS {\bf 110}, 6730 (2013).

\bibitem{HENNAN-KAMRIN PRL 2014} D. Hennan and K. Kamrin, Phys. Rev. Lett. {\bf 113}, 178001 (2014).


\bibitem{Fisher-etal-1997} D. Fisher, K. Dahmen, S. Ramanathan and Y. Ben-Zion, Phys. Rev. Lett. {\bf 78}, 4885 (1997).

\bibitem{EARTHQUAKES-RMP-1994} J.M. Carlson, J.S. Langer and B. Shaw,  Rev. Mod. Phys. {\bf 66}, 657 (1994).



\end{thebibliography}
\end{document}